\documentclass[a4, useAMS,usenatbib]{mn2e}

\pdfoutput=1
\usepackage[pdftex]{graphicx}
\usepackage[pdftex]{color}
\usepackage[colorlinks,bookmarks]{hyperref}
\definecolor{linkblue}{rgb}{0,0,0.8}
\definecolor{linkgreen}{rgb}{0,0.5,0}
\hypersetup{linkcolor=linkblue, citecolor=linkgreen, urlcolor=linkblue}

\usepackage[usenames,dvipsnames]{xcolor}
\usepackage[T1]{fontenc}
\usepackage[utf8]{inputenc}
\usepackage{lmodern}
\usepackage[english]{babel}
\usepackage{bm}
\usepackage{verbatim}
\usepackage[us]{datetime}
\usepackage{amsmath, amssymb}
\usepackage{tabularx,ragged2e,booktabs,caption}

\providecommand{\adsurl}[1]{\href{#1}{ADS}}

\def\rd{{\rm d}}

\newcommand{\post}{\mathfrak L}

\newcommand{\like}{\mathcal L}

\newcommand{\evi}{\mathcal E}
\newcommand{\prior}{\mathcal P}

\newcommand{\brat}{\mathcal B}

\newcommand{\thetabm}{\boldsymbol \theta}

\title[Extensive search for bias in SNIa data]{Extensive search for systematic bias in SN Ia data}

\author[C.~Heneka, V.~Marra and L.~Amendola]{Caroline Heneka$^{1,2}$, Valerio Marra$^{2}$ and Luca Amendola$^{2}$\\
$^{1}$Dark Cosmology Center, Niels Bohr Institute, University of Copenhagen, Juliane Maries Vej 30, 2100 Copenhagen, Denmark\\
$^{2}$Institut für Theoretische Physik, Universität Heidelberg, Philosophenweg 16, 69120 Heidelberg, Germany}

\begin{document}

\date{Accepted XXX. Received XXX; in original form XXX}

\pagerange{\pageref{firstpage}--\pageref{lastpage}} \pubyear{2014}

\maketitle

\label{firstpage}

\begin{abstract}
The use of advanced statistical analysis tools is crucial in order to improve cosmological parameter estimates via removal of systematic errors
and identification of previously unaccounted for cosmological signals. 
Here we demonstrate the application of a new fully-Bayesian method, the internal robustness formalism,
to scan for systematics and new signals in the recent supernova Ia Union compilations.
Our analysis is tailored to maximize chances of detecting the anomalous subsets
by means of a variety of sorting algorithms. 
We  analyse supernova Ia distance moduli for effects depending on angular separation, redshift, surveys and hemispherical directions.
The data have proven to be robust within $2\sigma$, giving an independent confirmation of successful removal of systematics-contaminated supernovae.
Hints of new cosmology, as for example the anisotropies  reported by Planck, do not seem to be reflected in the supernova Ia data.

\end{abstract}

\begin{keywords}
methods: statistical -- cosmology: cosmological parameters -- stars: supernovae: general
\end{keywords}

\section{Introduction} \label{intro}
The big quest in cosmology today is to put on firm grounds our understanding of cosmic acceleration, first discovered  by \citet{Perlmutter:1998np,Riess:1998cb}  with supernova Ia (SNIa) data. 
The presence of cosmic acceleration has since then been verified using
a number of SNIa datasets \citep{Hamuy:1996ss,Schmidt:1998ys,Perlmutter:1998np,Riess:1998cb,Knop:2003iy,Tonry:2003zg,Barris:2003dq,Krisciunas:2005dw,Jha:2005jg,Astier:2005qq,Miknaitis:2007jd,Riess:2006fw,Amanullah:2007yv,Kowalski:2008ez,Holtzman:2008zz,Hicken:2009dk,Contreras:2009nt}, 
now compiled together in the Union 2.0 and 2.1 catalogues \citep{Amanullah:2010vv,Suzuki:2011hu}.
A variety of other cosmological probes, e.g.~Baryonic Acoustic Oscillations (BAO) \citep{Eisenstein:2005su,Blake:2011wn}
and anisotropies of the Cosmic Microwave Background (CMB) \citep{Komatsu:2010fb,Ade:2013nlj,Aghanim:2013suk} confirm cosmic acceleration.
Especially now that we are entering an era of precision cosmology,
with the number of observed supernovae increasing significantly over the next 5 to 15 years by up to 1 or 2 orders of magnitudes -- for example with the Dark Energy Survey \citep{Bernstein:2009ue} and the Large Synaptic Survey Telescope \citep{Abell:2009aa} -- 
improvements of cosmological parameter estimation rely more and more on a better handling of our systematic error budget.

On the other hand, we strive as well to expand the interpretation of our results by revealing possible new cosmological signals
that have not been considered in a standard cosmological treatment of the data.
As cosmological parameter estimates and model comparisons can only be performed in a robust statistical framework,
especially given our situation of being unable to rely on controlled laboratory conditions,
we need to apply improved statistical tools to identify systematics or new cosmological signals
that are as yet unaccounted for. In other words, we naturally want to get as much as we can out of the data available.
But how can this be done?

Here we want to focus on analyses of SNIa data, 
more specifically the recent Union 2.0 and 2.1 catalogues \citep{Amanullah:2010vv,Suzuki:2011hu} that have been compiled from a range of different surveys, 
taking into account different possible systematics and strategies to appropriately standardize the supernovae;
it should be stressed though that also other types of data can be tested via the method outlined here.
Some of the known effects that could alter the SNIa apparent magnitudes are
local deviations from the Hubble flow \citep[as e.g.~in][]{Marra:2013rba}, dust absorption \citep{Corasaniti:2006cv,Menard:2009wy}, lensing by foreground structures \citep{Jonsson:2010wx,Amendola:2013twa, Quartin:2013moa}, a change of systematics when moving between observational bands
or supernovae coming from different populations \citep[see e.g.][]{Astier:2005qq,Wang:2013jpa}.
Additional cosmological effects altering the supernova magnitudes can be described by non-standard models,
such as inhomogeneous models displaying a variation of the expansion rate with redshift \citep[see the review][and references therein]{Marra:2011ct},
or anisotropic models with anisotropic expansion rates as in \citet{Graham:2010hh}.

A wide range of statistical tests and cross-checks is already being applied to the data 
so as to assess the ability of a model to describe observations 
(e.g.~goodness-of-fit test, which however is not sensitive to the full likelihood),
as well as to compare the performance of different models (e.g.~likelihood-ratio test).
However, analyses performed so far of these effects have always assumed a specific type of effect to then estimate its statistical significance.
A previously introduced fully Bayesian method, the BEAMS formalism \citep{Kunz:2006ik}, estimates parameters based on the probability of data belonging to different underlying probability distributions, thus dealing with different underlying populations present, and includes the treatment of correlations in \citet{Newling:2011cp}.

As we do want to be as unprejudiced as possible and do not want to speculate about the exact nature of possible deviations from the overall model estimate, 
we will use a model-independent tool---the fully Bayesian method introduced in~\cite{Amendola:2012wc}, dubbed \emph{internal robustness}.
This method searches for statistically significant signals of incompatible subsets in the data, 
without assuming any specific model and taking into account the full likelihood when forming Bayesian evidences.
The internal robustness is able to identify subsets of supernovae that can be better described by a set of parameters differing from the best-fit model of the overall set, i.e.~we do not search for single outliers but instead search for incompatible sub-populations in the data.
In a more abstract statistical sense, we search for subgroups having a deviating trend in the variance, 
a property that is sometimes called heteroscedasticity.

Another particularity here is that, in addition to blind analysis,
we want to raise our chances of finding the subsets that are most likely to be biased by applying a suite of sorting algorithms to the data.
In principle there exists a variety of ways to partition the SNIa data,
for example sorting by angular separation between pairs of supernovae, 
motivated by the suspicion that angularly clustered supernovae undergo comparable systematic effects,
or to focus on new cosmological signals by e.g.~testing the isotropy of the data.
Our  goal is to assess the robustness of  SNIa data with regard to systematics or 
hints of unaccounted for cosmological signals, 
and to identify systematically biased subsets in order to improve cosmological parameter estimation.

In Section~\ref{sec:form} we recapitulate the formalism introduced in~\cite{Amendola:2012wc} and introduce its extension to systematic parameters.
Section~\ref{sec:methods} describes the real and synthetic catalogues used and the internal robustness calculation procedure.
Section~\ref{sec:results} presents the analysis and results for the robustness test of the Union 2.0 and 2.1 catalogues in a
angular seperation-, redshift-, survey- and directional-dependent way to look for systematics or new signals of inhomogeneity or anisotropy.
We will discuss our findings in Section~\ref{sec:conco}.

\section{Formalism}
\label{sec:form}

\subsection{Bayesian evidence and internal robustness}\label{sec:robust}

Bayes' theorem allows to obtain 
the conditional probability $\post(\thetabm^{M};\bm{x} )$ 
of the $n$ theoretical parameters that describe the model $M$,
$\thetabm^{M} = (\theta_{1},...,\theta_{n})$, 
given the $N$ random data ${\bm{x}}=(x_{1},...,x_{N})$.
It states~\citep[see e.g.][]{trotta}:
\begin{equation}
 \post(\thetabm^{M};\bm{x} )=\frac{ \like ({\bm{x}};\thetabm^{M}) \, \prior(\thetabm^{M}) }{ \evi({\bm{x}};M)} \,, 
\label{bayes}
\end{equation}
where $\like ({\bm{x}};\thetabm^{M}) $  is the likelihood of having the data $\bm{x}$ given the model parameters $\thetabm^{M}$, $\prior(\thetabm^{M})$ is 
the prior on the parameters
and $ \evi({\bm{x}};M)$ is the normalization. 
The normalization is often referred to as Bayes evidence and can be calculated~via
\begin{equation}
    \evi({\bm{x}};M)=\int \like ({\bm{x}};\thetabm^{M}) \prior(\thetabm^{M})\,\rd^{n}\thetabm^{M}\,.
\label{eq:likebay}
\end{equation}
Applying Bayes' theorem a second time, one obtains the posterior probability $\post(M;{\bm{x}})$ of model $M$ under data $\bm{x}$:
\begin{equation}
    \post(M;{\bm{x}})=\evi({\bm{x}};M)\frac{\prior(M)}{\prior({\bm{x}})}\,,
\end{equation}
where $\prior(M)$ is the prior on a particular model $M$ and $\prior({\bm{x}})$ is the (unknown) probability of having the data $\bm{x}$.
We can then compare quantitatively the performance of two models $M_{1}$ and $M_{2}$ to describe the data by 
taking the ratio of the posterior probabilities ($\prior({\bm{x}})$ cancels out):
\begin{equation}
    \frac{\post(M_{1};{\bm{x}})}{\post(M_{2};{\bm{x}})}=\brat_{12}\frac{\prior(M_{1})}{\prior(M_{2})}\,,
\end{equation}
with the Bayes ratio $\brat_{12}$ being 
\begin{equation}
    \brat_{12}
    = \frac{\evi({\bm{x}};M_{1})} {\evi({\bm{x}};M_{2})}
    \,.\label{eq:bayesratio}
\end{equation}
It is usually assumed that $\prior(M_{1})=\prior(M_{2})$ so that the Bayes ratio $\brat_{12}>1$ says that current data favors the model $M_{1}$, and \emph{vice versa}.

To come back to our aim of testing the robustness of SNIa data,
we compare two alternative hypotheses concerning the underlying models, following the formalism introduced
 in~\cite{Amendola:2012wc}, which extends the  previous results of~\cite{2011MNRAS.415..143M}.
The first hypothesis is  that all  data ($\textbf{x}_{\rm tot}$) is best described by one overall 
model $M_C$; the alternative hypothesis is that
  data is composed of two (complementary) subsets -- $\textbf{x}_{1}$ and $\textbf{x}_{2}$  -- which are  described by two independent models, $M_C$ and $M_S$ respectively.
The first model is referred to as the ``cosmological'' model, while the second one as the ``systematic'' model, which is a model, other than $M_C$, that well describes a subset of the data. This could be due to the fact that
part of the dataset is heavily affected by experimental errors or because intrinsically they are different, e.g.~supernovae with different progenitors.
The statistical significance of the preference for one of these assumptions is assessed
by comparing the correspondent Bayesian evidences:
\begin{equation}
\brat_{\rm tot,ind}
=\frac{\evi_{\rm tot}}{\evi_{\rm ind}}
=\frac{\evi\left(\textbf{x}_{\rm tot};M_{C}\right)}{\evi\left(\textbf{x}_{1};M_{C}\right)\evi\left(\textbf{x}_{2};M_{S}\right)} \,,
\label{eq:Bcomb}
\end{equation}
where the evidence for the independent model assumption is simply the product of the individual evidences.
The logarithm of the Bayes' ratio~(\ref{eq:Bcomb}),
\begin{equation}
R\equiv\log \brat_{\rm tot,ind},
\label{eq:rob}
\end{equation}
dubbed \emph{internal robustness}, is now a suitable quantity to test the assumption of having one underlying model
instead of two independent ones. 
This search will be conducted by integrating the evidences via~(\ref{eq:likebay})
and calculating the corresponding $R$ for the chosen partitions $\textbf{x}_{1,2}$ of the dataset.

If the subset sizes are sufficiently big, the Fisher approximation can be used and the likelihood functions can be approximated as Gaussian both in  data and in  parameters.
In Section 3 of \cite{Amendola:2012wc} it was empirically found that the Fisher approximation can be used if the smaller subset has more than $N_{\rm min} = 90$ elements.
The evidence of the (very large) complementary set is always computed using the Fisher approximation.
In this paper, the Fisher-approximated internal robustness was only used for the robustness calculations of Section \ref{hemis}, where subset sizes are well above $N = 100$.
The Fisher-approximated internal robustness, as derived in~\cite{Amendola:2012wc}, is then:
\begin{equation}
R = R_{0} -\frac{1}{2}\left( \widehat \chi_{\rm tot}^2 - \widehat \chi_{1}^2 - \widehat \chi_{2}^2 \right) + \frac{1}{2} \log\left( \frac{|L_{1}| |L_{2}|}{|L_{\rm tot}|} \right),
\label{eq:robfish}
\end{equation}
where correlations have been neglected.
$R_{0}$ includes the unknown systematic prior determinant, 
and the quantities $\widehat \chi_{\rm tot}^2$, $\widehat \chi_{1}^2$ and $\widehat \chi_{2}^2$ are the bestfit chi-square values for the overall set, subset 1 and complementary subset 2, respectively.
The third term takes into account the change in parameter space volume
via the ratio of the determinants of the Fisher matrices for the set $\textbf{x}_{\rm tot}, \textbf{x}_{1,2}$.

\subsection{Cosmological parametrization} \label{sec:fisher}

In our analysis the observable is the apparent magnitudes of the supernovae. The likelihood for the case of the cosmological parametrization -- marginalized over  absolute magnitude and present-day value of the Hubble rate $H_{0}$ 
 -- is as usual~\citep{Amendola:2012wc}:
 \begin{equation}
-\log \like = \sum_{i}^{N'}\log({\sqrt{2\pi}\sigma_{i}} )+ \frac{1}{2}\log{\frac{S_{0}}{2\pi}} + \frac{1}{2}\left( S_{2} - \frac{S_{1}^2}{S_{0}} \right),
\label{eq:Lmarg}
\end{equation}
where we neglected correlations among supernovae and $N'$ denotes the number of elements in the dataset.
The sums $S_{n}$ are defined as
\begin{equation}
S_{n}=\sum_{i}^{N'}\frac{\delta m_{i}^n}{\sigma_{i}^2} \,,
\label{eq:sums}
\end{equation}
where $\delta m_{i}=m_{{\rm obs,i}}-m_{{\rm th,i}}$ are the magnitude residuals, i.e.~the differences between observed apparent magnitudes and theoretically expected ones.

The Fisher matrix in terms of $S_{n}$ and  derivatives is
 \begin{equation}
F_{pq} \equiv -\frac{\partial^{2}\,\log \like }{\partial\theta_{p}\partial\theta_{q}} = {1 \over 2} S_{2,pq} - \frac{1}{S_{0}}\left( S_{1}S_{1,pq} + S_{1,p}S_{1,q}\right) ,
\label{eq:fishS}
\end{equation}
where the comma denotes derivative with respect to model parameters. 
\newline In the cosmological parametrization the predicted magnitude is calculated via the cosmology-dependent luminosity distance $d_{L}$:
\begin{equation}
m_{\rm th,i} (z)=5\log_{10}  d_{L} (z_{i}) \,,
\end{equation}
where  $d_{L}$ is in units of the (irrelevant) $H_0^{-1}$.
From Eq.~(\ref{eq:fishS}) it follows then:
 \begin{align}
F_{pq}= & \frac{5}{\ln10}\sum_{i}{\frac{1}{\sigma_{i}^2}}\left(\frac{d_{L i,p}d_{L i,q}}{d_{L i}^2}-\frac{d_{L i,pq}}{d_{L i}} \right) \left(\delta m_{i}-\frac{S_{1}}{S_{0}}\right)\nonumber \\
+ & \frac{25}{(\ln10)^2}\left(\sum{\frac{d_{L i,p}d_{L i,q}}{\sigma_{i}^2d_{L i}^2}}-\frac{1}{S_{0}}\sum_{i}{\frac{d_{L i,p}}{\sigma_{i}^2d_{L i}}}\sum_{j}{\frac{d_{L j,q}}{\sigma_{j}^2 d_{L j}}}\right).
\end{align}

As pointed out earlier, in the present analysis we are neglecting correlations in the distance moduli of the supernovae (a possible correlation between the errors is inconsequential).
Correlations stem from the fact that we will use processed rather than raw data, so as to simplify the numerically challenging task of obtaining the evidence.
This caveat should be kept in mind when interpreting our findings as it may potentially decrease the sensitivity of the internal robustness test.

\subsection{Systematic parametrization}\label{sec:fisher2}

The parameters that describe the systematic model are in general different from the ones that describe the overall cosmological model. We adopt here two opposite philosophies. In one
we test the hypothesis that a data subset is described by a different cosmology, still parametrized by the same cosmological parameters of the overall model for $\textbf{x}_{\rm tot}$, e.g. $\Omega_m,\Omega_{\Lambda}$. This is in some cases the obvious choice, for instance when we test the idea that the universe
is anisotropic and therefore the cosmological parameters in one direction are different from those in another.

The second philosophy is that if we have no clue of what the $M_S$ parameters could be then we can just make the simplest choice, i.e.~a linear model.
In this second case  the phenomenological parametrization
 can be chosen as:
 \begin{equation}
m(z)=\sum_{i}{\lambda_{i} \,f_{i}(z)} \,,
\end{equation}
with parameters $\lambda_{i}$ and the redshift-dependent functions $f_{i}(z)$. 
The parametrized observable does not necessarily have to be the magnitude,
it can be any other observable we choose to analyse.
We take $f_{i}(z)$ to be polynomials in the redshift $z$, so that the observable parametrized by $n$ parameters is
 \begin{equation}
m(z)=\sum_{i=0}^{n}{\lambda_{i} \, z^{i}} \,.
\label{eq:phenparam}
\end{equation}
As this parametrization is linear in the $\lambda_{i}$, the second derivatives in Eq.~(\ref{eq:fishS}) vanish 
and the Fisher matrix becomes
 \begin{align}
F_{pq}= 
\sum_{i}{\frac{f_{i,p}f_{i,q}}{\sigma_{i}^2}}  
- \frac{1}{S_{0}}\sum_{i}{\frac{f_{i,p}}{\sigma_{i}^2}}\sum_{j}{\frac{f_{j,q}}{\sigma_{j}^2}}.
\end{align}
For the systematic parametrization the best-fit is analytical as well
and can be found easily by maximizing the likelihood.
By making use of Eq.~(\ref{eq:Lmarg}) one finds:
 \begin{equation}
\left.\frac{1}{2}S_{2,q}-\frac{S_{1}S_{1,q}}{S_{0}}\right|_{\lambda=\lambda_{p}}=0.
\end{equation}
Inserting the sums~(\ref{eq:sums}) and replacing the parametrized residuals $\delta m _{i}= m_{\rm obs,i}-\lambda_{j}f_{j}(z_{i})$ gives
\begin{align}
\sum_{i} & \frac{m_{\rm obs,i}f_{i,q}}{\sigma_{i}^2}-\frac{1}{S_{0}}\sum_{i}\frac{m_{\rm obs,i}}{\sigma_{i}^2}\sum_{j}\frac{f_{j,q}}{\sigma_{j}^2} \nonumber \\
- & \left.\sum_{i}\frac{\lambda_{k}f_{k}f_{i,q}}{\sigma_{i}^2}+\frac{1}{S_{0}}\sum_{i}\frac{\lambda_{k}f_{k}}{\sigma_{i}^2}\sum_{j}\frac{f_{j,q}}{\sigma_{j}^2}\right|_{\lambda=\lambda_{p}}=0 \,.
\end{align}
The best-fit parameters $\lambda_{p}$ can then be calculated via
\begin{equation}
\lambda_{p}=F_{pq}^{-1}\left(\sum_{i} \frac{m_{\rm obs,i}f_{i,q}}{\sigma_{i}^2}-\frac{1}{S_{0}}\sum_{i}\frac{m_{\rm obs,i}}{\sigma_{i}^2}\sum_{j}\frac{f_{j,q}}{\sigma_{j}^2}\right).
\end{equation}
We make use of this phenomenological parametrization 
either when cosmological parameter estimation fails, for example due to subset sizes being too small,
or when searching for a purely systematical signal in the data.

\section{Methodology}
\label{sec:methods}

\subsection{Real catalogues}
The data used for our analyses are the supernova Union2.0 compilation~\citep{Amanullah:2010vv} of 557 supernovae with  redshifts ranging from
$z=0.015$ to $z=1.4$ and the updated Union2.1 compilation~\citep{Suzuki:2011hu} of 580 supernovae with redshift in the range from $z=0.015$ to $z=1.414$.
The Union2.1 compilation adds to the 17 surveys compiled together in Union2.0 two more recent surveys, while discarding due to new quality cuts some previously included supernovae. We choose these two compilations due to their widespread employment in cosmological parameter inference and the wide range of redshift and partial surveys they cover. 
Throughout the paper our observable is apparent magnitudes, stretch and color corrected. We used global stretch and color correction parameters, which have been fixed to the best-fit values: $\{ \alpha, \beta \} = \{0.1219, 2.466 \}$ for the Union2.1 compilation and $\{ \alpha, \beta \} = \{0.1209, 2.514 \}$ for the Union2.0 compilation.
Consequently, a possible redshift dependence of the color parameter $\beta$ \citep[see][]{Kessler:2009ys} has been neglected.

\subsection{Creation of synthetic catalogues}
As no analytical form for the 
 expected internal robustness probability distribution function (iR-PDF) is available for now,
unbiased synthetic catalogues have to be created to test for the significance of the internal robustness values obtained for the real catalogue.
The iR-PDF is indeed a very non-trivial object, formed by sampling possible partitions within a fixed overall realization~\citep[see][Section 2.3]{Amendola:2012wc}.
The synthetic catalogues were created by adding  a  Gaussian error to the best-fit function of the distance modulus, using as $\sigma$ the distance modulus errors of the real catalog.

\subsection{Creation of sublists} \label{sec:create}
As it is not feasible to scan all possible partitions of the Union catalogues,
different strategies for partitioning the data have to be employed in order to test the subsets 
and their underlying best-fit model parameterizations for their compatibility with the complementary subsets.
One possibility is to randomly pick a number of subsets out of all possible subsets, 
constraining the subset size to vary between some minimal value necessary to determine the model parameters
and a maximum of half the size of the total set.
This way of partitioning the data is chosen when one does not want to test for a specific prejudice,
but instead to search for any possible signal and has been carried out in \cite{Amendola:2012wc}.
There it was found that the Union 2.1 Compilation does not possess a significant amount of systematics.

To test a certain prejudice  regarding the occurrence of either systematics or new cosmological signals,
we divide the total set into subsets in a way to maximize the chances of finding a subset of low robustness.
This approach has a potential sensitivity higher than the one relative to the blind search carried out in \cite{Amendola:2012wc}.
We partition SN data according to the following criteria:

\begin{itemize}

\item
Section \ref{anguse}: subsets  chosen according to angular separation on the sky,

\item
Section \ref{hemis}: data  divided into  hemispheres,

\item
Section \ref{cosmoz}: data  partitioned according to redshift,

\item
Section \ref{surv}: supernovae  grouped according to their survey of origin.

\end{itemize}
It should be noted that while we select partitions according to a given prejudice, the statistics that we use -- the internal robustness test -- remains unchanged. This should make our analysis robust and fair, avoiding the risk of using a statistics which has potentially been tailored \emph{a posteriori}.

\subsection{Robustness analysis} \label{ana}
The analysis of the Union catalogue is conducted as follows.
The internal robustness is calculated following the formalism introduced by~\cite{Amendola:2012wc} and briefly summarized in Section~\ref{sec:robust}.
To do so, the observables were parametrized either cosmologically or phenomenologically,
as discussed in Sections~\ref{sec:fisher}-\ref{sec:fisher2}.
After having chosen a way to partition the data,
the robustness value for each chosen partition was calculated
 for real as well as for unbiased synthetic catalogues.
For a set of partitions one thus obtains an internal robustness probability distribution function (iR-PDF).
The iR-PDF of the real catalogue  is then to be compared to the iR-PDF of the synthetic catalogues
in order to assess the significance of the signal, as an analytical form for the iR-PDF is not available.
Possible deviations between real iR-PDF and synthetic unbiased iR-PDF tell us
 how compatible are the sublists formed with each other or rather their underlying best-fit models.
A strong incompatibility of robustness values between real catalogue and synthetic unbiased catalogues therefore is a signal 
for possible unaccounted for systematics or new cosmological signals that influence the cosmological parameter estimation.


\section{Results}
\label{sec:results}

\subsection{Angular separation}       
\label{anguse}

\begin{figure}
\begin{center}
\includegraphics[width= \columnwidth]{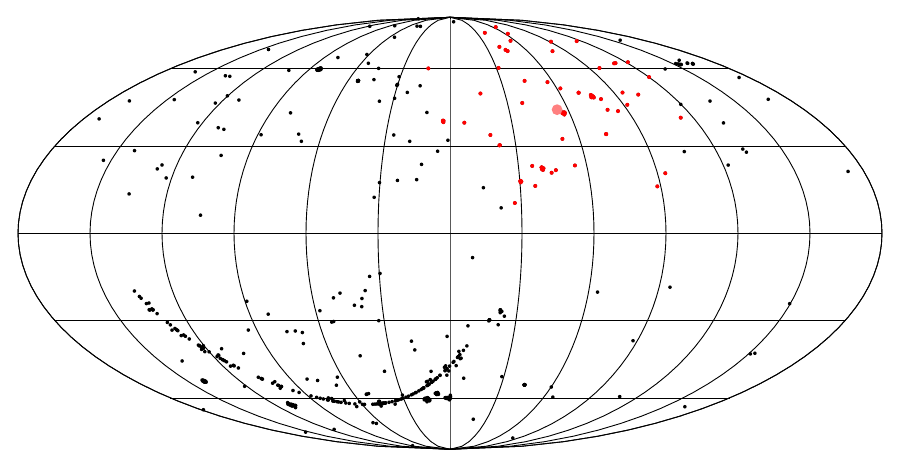}
 \qquad
\caption{Mollweide plot of the Union2.0 Compilation~\citep{Amanullah:2010vv}.
Red dots show the 80 nearest (in terms of angular separation) supernovae to the supernova marked with a larger and lighter dot.
Black dots show the complementary set.}
\label{FIG:mollsep1}
\end{center}
\end{figure}

In this section we analyse the robustness of supernovae sorted by angular separation $\Delta\theta$ on the celestial sphere, which can be found using the following relation: 
\begin{align}
\cos(\Delta\theta) = &  \sin(90^{\circ}+\delta_{1})\sin(90^{\circ}+\delta_{2})\cos(\alpha_{1}-\alpha_{2}) \nonumber \\  
+ & \cos(90^{\circ}+\delta_{1})\cos(90^{\circ}+\delta_{2}) \,,
\label{eq:angsep}
\end{align}
where $\alpha$ and $\delta$ are right ascension and declination, respectively.
We will use a cosmological parametrization (see Section \ref{sec:fisher}) for their distance moduli.
This angular sorting is expected to maximize our chances of finding a signal due to e.g.~dust extinction affecting angularly grouped supernovae.

\subsubsection{Fixed subset sizes} \label{sec:cosmosep}

We will carry out our analysis in two steps. 
In the first step, for each supernova of the Union2.0 Compilation we form a subset made of the 10-to-80 nearest supernovae, for a total of $71 \cdot 557 = 39547$ subsets.
The angular extension of the subsets is not constant (we will carry out a complementary analysis in Section \ref{sec:cosmosep5deg}), and it will be larger when the central supernova belongs to a region of the sky with few supernovae or smaller when belonging to a dense region, such as the SDSS stripe (clearly visible in the lower left of Fig.~\ref{FIG:mollsep1}).
It is also interesting to point out that subsets in dense regions are more likely to contain supernovae at more similar redshifts than supernovae in sparse regions.
An upper bound of 80 for the size of the smaller subset of the partition was chosen so as to cover a not too large area of the sky.
A typical partition is illustrated in Fig.~\ref{FIG:mollsep1}.
Partitions with larger fractions of the sky will be covered by the hemispherical analysis of Section \ref{hemis}.
The lower bound on the subset size of 10 supernovae was chosen such that the percentage of subsets,
for which the procedure of model parameter estimation and robustness calculation fails, 
makes up around one percent at most of the total subset population.

\begin{figure*}
\includegraphics[width=2 \columnwidth]{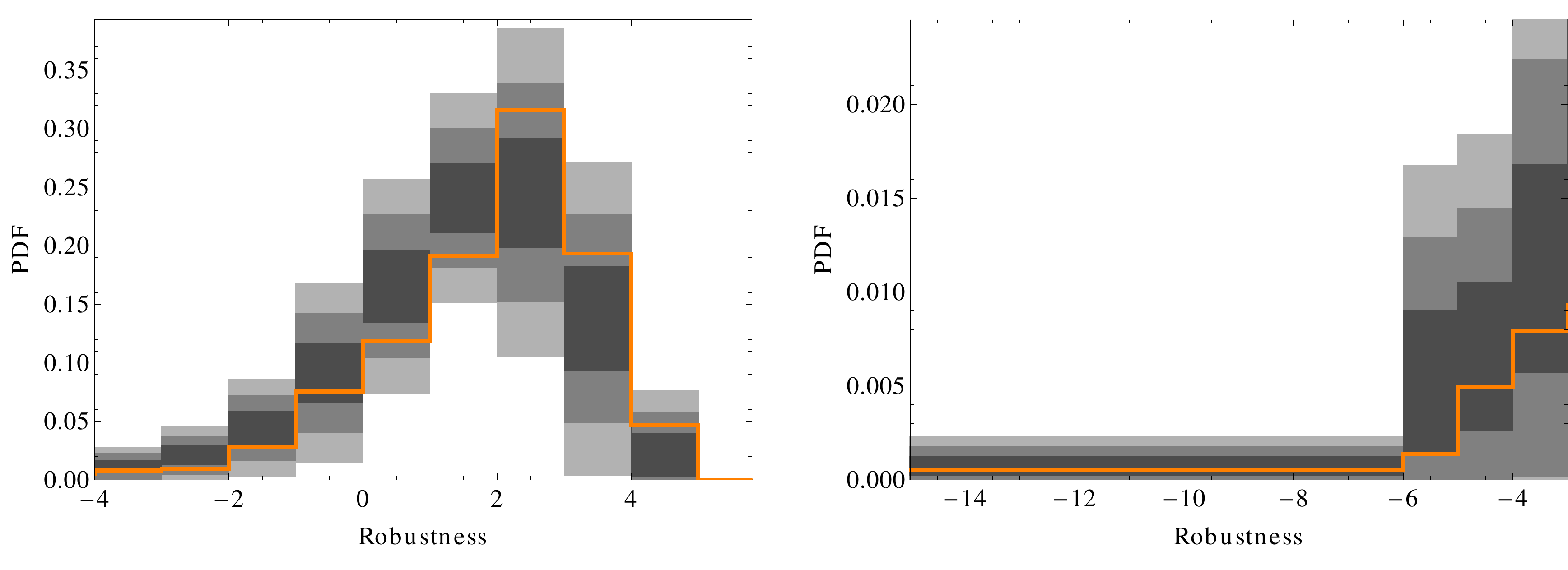}
\caption{Left panel: Binned iR-PDF of Union2.0 Compilation (orange solid line) obtained by sampling partitions according to their angular separation.
More precisely, for each supernova a subset made of the 10-to-80 nearest supernovae was formed.
In grey Gaussian 1, 2, 3-$\sigma$-bands from 145 unbiased synthetic catalogues are shown.
Right panel: zoom on the low-robustness tail.
As can be seen, the Union2.0 PDF is always within the 2-$\sigma$-band, and we can conclude that the catalogue seems robust with regards to systematics possibly related to the angular position of supernovae.
See Section \ref{sec:cosmosep} for more details.
}
\label{FIG:mockr1}
\end{figure*}

\begin{figure*}
\includegraphics[width=2 \columnwidth]{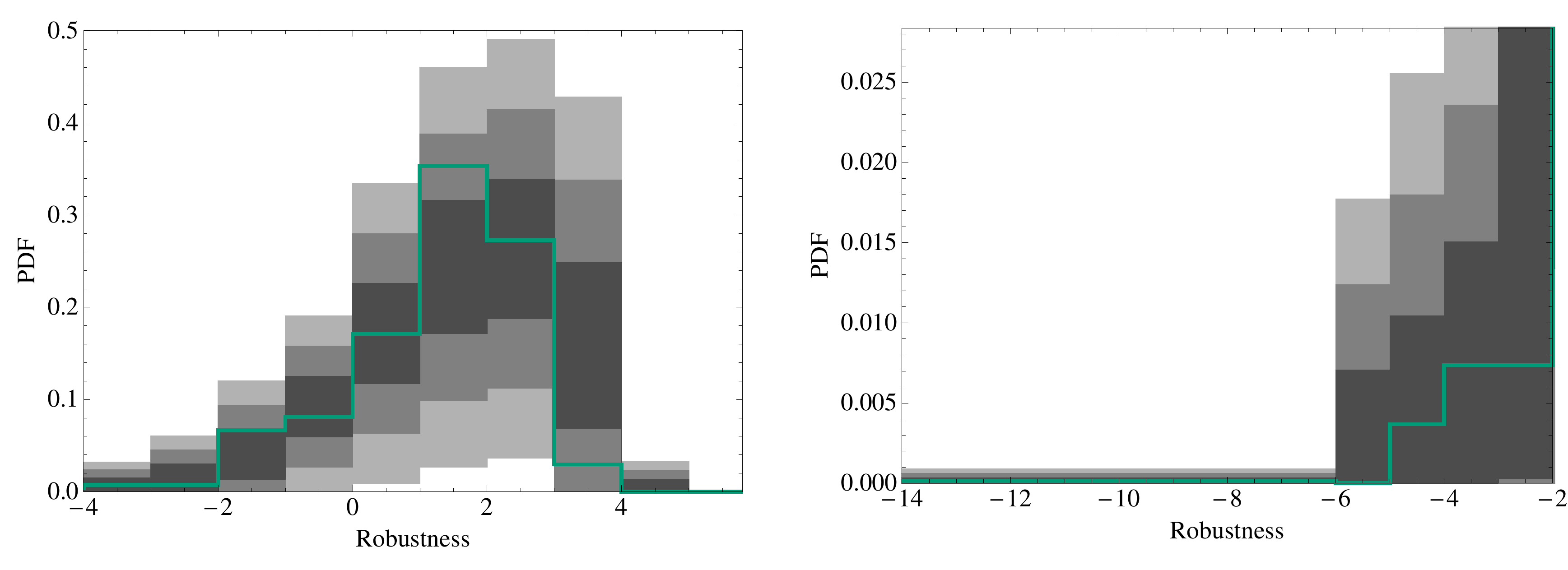}
\caption{
Same as in Fig.~\ref{FIG:mockr1}, but for subsets formed by grouping all supernovae within $5^\circ$ of a given supernova in the Union2.1 catalog.
As can be seen, the Union2.1 PDF always lies within the 2-$\sigma$-band and so the catalogue seems robust with regards to systematics possibly related to the angular position of supernovae.
See Section \ref{sec:cosmosep5deg} for more details.
}
\label{FIG:mock5deg1}
\end{figure*}

Small supernova datasets have indeed likelihoods which tend to be partially degenerate and spread on large supports. It is numerically problematic to compute the evidence for subsets with less than 10 supernovae -- even in the case of the very extend parameter space of~(\ref{parL}) -- and we therefore exclude such partitions from our analysis.
We use the following parameter space for the cosmological parameters $(\Omega_{m},\Omega_{\Lambda})$:
\begin{equation}
    -20\leq\Omega_{m}\leq20
  \quad  \textrm{and} \quad
    -45\leq\Omega_{\Lambda}\leq20 \,,
\label{parL}
\end{equation}
which is much broader than the conventional physical one as it is supposed to also describe possible systematic effects.
Following the approach described above, we were able to successfully compute robustness values for 38858 subsets, from which the binned iR-PDF of the Union2.0 catalog was  obtained (orange solid line in Fig.~\ref{FIG:mockr1}). Our computing scheme failed for only about one per cent of the subsets, thus achieving the desired performance goal.

The procedure for the obtaining the binned iR-PDF has been repeated -- in exactly the same way -- for 145 unbiased synthetic catalogues.
The distribution of synthetic iR-PDF in a given robustness bin allows then to assess the significance of possible anomalous signals in the iR-PDF of the Union2.0 catalogue.
In Fig.~\ref{FIG:mockr1}, we show in grey Gaussian 1, 2, 3-$\sigma$-bands and the mean of the synthetic catalogues, together with the binned iR-PDF for the real catalogue in solid orange.
As can be seen, the Union2.0 PDF is always within the 2-$\sigma$-band, and we can conclude that the catalogue seems robust with regards to systematics, 
even when having limited the analysis to angular separation-sorted sublists.

\subsubsection{Fixed angular separation}
\label{sec:cosmosep5deg}

In the second step, we want to keep fixed the angular scale of the subsets tested.
For each supernova of the Union2.1 Compilation, a subset is formed by selecting all the supernovae within an angular separation of $5^\circ$, amounting to subset sizes ranging from 10 to 62 supernovae. This amounts to 351 subsets tested, as subsets containing less than 10 supernovae were removed in order to ensure parameter estimation.  We chose this angular scale in order to test small areas on the sky for possible deviations of their properties with respect to the full sky.
This analysis is therefore complementary to the one of the previous Section \ref{sec:cosmosep}.

In Fig.~\ref{FIG:mock5deg1} we show the results for the chosen angular separation of $5^\circ$.
Again, no signal of systematics was found.

\subsection{Hemispherical anisotropy} \label{hemis}

\begin{table}
\caption{Significance in $\sigma$-units of the robustness value of the Union2.1 Compilation with respect to unbiased synthetic catalogues for various directions of hemispherical anisotropy.
See Section \ref{hemis} for more details.}
\centering
\begin{tabular}{l c r}
\hline \hline
Type &  $(\alpha,\delta)$ & Significance  \\
\hline
\begin{minipage}[c][.73cm][c]{4 cm}
\begin{flushleft}
Hemispherical asymmetry \citep{Ade:2013nlj}
\end{flushleft}
\end{minipage} & $(270^{\circ},66.6^{\circ})$ &  $1.26\sigma$ \\ 
\begin{minipage}[c][.73cm][c]{3cm}
\begin{flushleft}
Dipole anisotropy \citep{Aghanim:2013suk}
\end{flushleft}
\end{minipage} & $(167^{\circ},-7^{\circ})$ &  $0.39\sigma$ \\ 
\begin{minipage}[c][.73cm][c]{4 cm}
\begin{flushleft}
Quadrupole-octupole alignment \citep{Ade:2013nlj}
\end{flushleft}
\end{minipage}  & $(177.4^{\circ},18.7^{\circ})$ &  $0.35\sigma$ \\ 
\hline
\begin{minipage}[c][.73cm][c]{4 cm}
\begin{flushleft}
Direction of lowest robustness (see Section \ref{hemigri})
\end{flushleft}
\end{minipage} & $(150^{\circ},70^{\circ})$ &  $2.20\sigma$ \\
\hline \hline
\end{tabular}
\label{table:hemi}
\end{table}

In the previous Section \ref{anguse}, we searched for signals of low robustness by grouping supernovae according to their angular position. In particular, the idea was to search for small subsets of supernovae that -- if found systematics driven -- could be removed from the full dataset in order to improve parameter estimation.
In this Section we will perform a similar analysis by partitioning the dataset into hemispheres.
The aim, however, will not be to purge the dataset of systematics driven supernovae but rather to search for a cosmological signal suggesting large-scale anisotropies in the universe.

Indeed, signals suggesting deviation from isotropy have already been detected, using both SNIa data~\citep{Colin:2010ds,Kalus:2012zu,Cai:2013lja,Yang:2013gea,Rathaus:2013ut} and CMB maps~\citep{Ade:2013nlj,Aghanim:2013suk}. Depending on the analysis the anisotropic signal is more or less in agreement with the expected one in a $\Lambda$CDM universe. Therefore, further analyses are required so as to understand if there are or not reasons to suspect a departure from the standard model of cosmology.



\subsubsection{Hemispheres for special directions}

\label{hemispe}

We will search for hemispherical anisotropy following two approaches.
The first one consists in examining directions along which anisotropic signals have been found:
the direction of hemispherical asymmetry (quite coinciding with the ecliptic plane), the one of the one of dipole anisotropy and the one of quadrupole-octupole alignment (chosen as the quadrupole direction of maximal quadrupole-octupole alignment), as summarized in Table~\ref{table:hemi}.

\begin{figure*}
\begin{center}
\includegraphics[width= \textwidth]{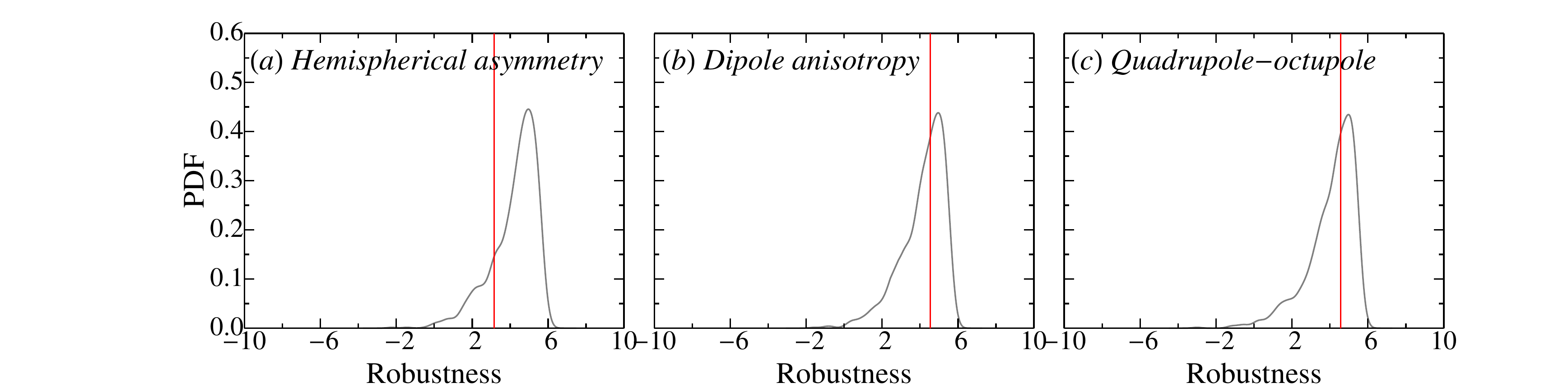}
\caption{
Tests for the three hemispherical directions report by the Planck Collaboration:
the hemispherical asymmetry (left), the dipole anisotropy (center) and the quadrupole-octupole alignment (right);
see Table~\ref{table:hemi} for the angular coordinates.
The red vertical lines show the internal robustness values of the Union2.1 Compilation, which are always well within the distribution of robustness values from the 1000 unbiased synthetic catalogues analysed.
See Section \ref{hemispe} for more details.
}
\label{FIG:hemiplanck}
\end{center}
\end{figure*}

In order to test for these three directions of anisotropy, 
the data were divided into hemispheres with their poles centered on the directions of maximal anisotropy.
This partitioning clearly yields one single robustness value for the Union2.1 catalogue.
To test for the significance of the results, 
we  perform the robustness analysis for 1000 synthetic catalogues partitioned into hemispheres in the same way as the real catalogue.
The parametrization chosen for the analysis is the standard cosmological one as we are looking for a signal of cosmological origin. Furthermore, this will help in comparing with previous results as the latter use the framework of standard cosmology.
We show the results of this analysis in Fig.~\ref{FIG:hemiplanck}.
As can be seen, the red vertical line -- corresponding to the Union2.1 Compilation -- is always well within the body of the distribution of robustness values from the synthetic catalogues.
Therefore, we conclude that the directions reported by the Planck Collaboration do not seem to be reflected,
at least not at a significant level, in supernova data.

In addition to the preferred Planck directions tested we find low significance and therefore low level of anisotropy for the Union2.1 preferred direction reported in~\cite{Yang:2013gea}.

\subsubsection{Grid of hemispheres}
\label{hemigri}

\begin{figure*}
\includegraphics[width=2 \columnwidth]{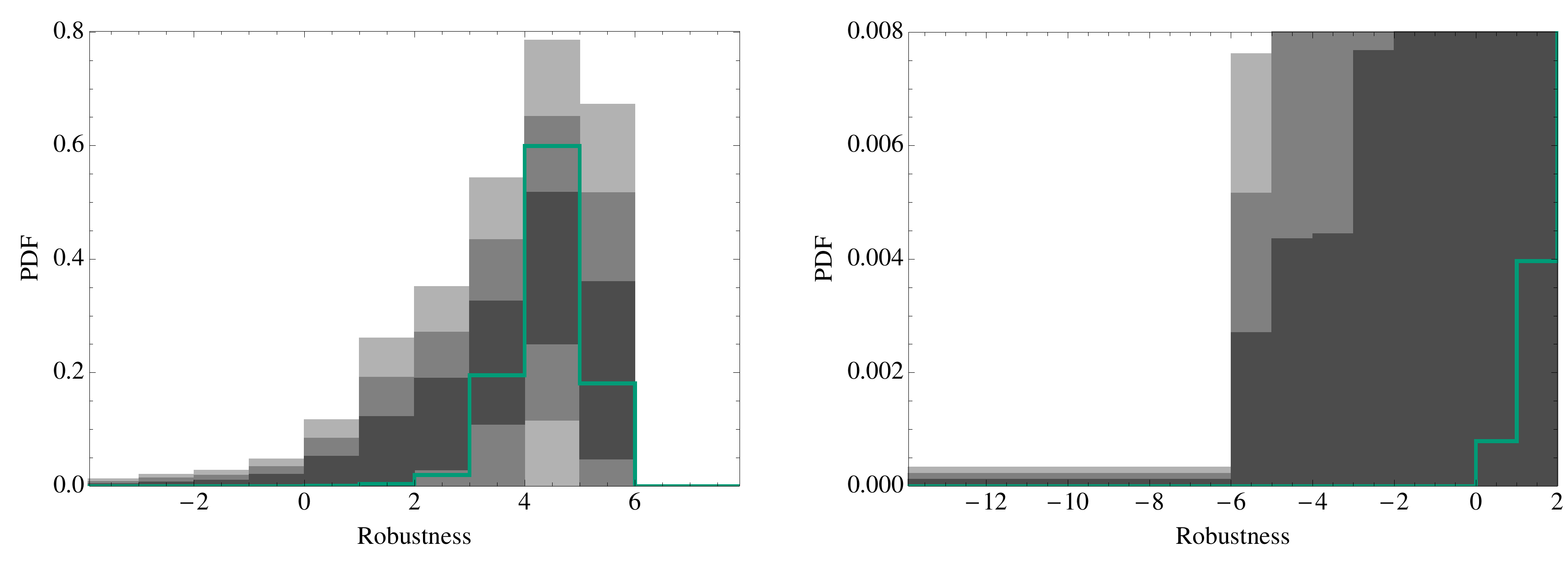}
\caption{
Same as in Fig.~\ref{FIG:mockr1}, but for subsets formed by grouping all supernovae in hemispheres whose north poles lie on a grid of $5^{\circ}$x $5^{\circ}$ on both spherical coordinates.
As can be seen, the Union2.1 PDF lies within the 2-$\sigma$-band and so the catalogue seems robust with regards to possible hemispherical anisotropy .
See Section \ref{hemigri} for more details.
 }
\label{FIG:hemigrid}
\end{figure*}

The second approach consists in testing a grid of hemispherical directions in order to determine the least robust one.
To do so, we drew a grid of $5^{\circ}$x $5^{\circ}$ on both spherical coordinates, 
whose intersections determine the directions of hemispherical poles.
The robustness values for the corresponding partitions was then computed.
The same procedure on the same grid was then followed for 100 synthetic catalogues.
The iR-PDFs of the real catalogue in green with $\sigma$-bands from the synthetic catalogues in grey is shown in Fig.~\ref{FIG:hemigrid}.
As can be seen, the real catalogue stays within the $2\sigma$-band: no significant direction-dependent effect can therefore be detected.
The hemispherical direction of lowest robustness can be found in  Table~\ref{table:hemi}, and does not point in a direction similar to any of the anisotropic directions reported by the Planck Collaboration.
In order to find the significance of this specific direction, we followed the procedure of the previous Section~\ref{hemispe}.

\subsection{Redshift dependence} \label{cosmoz}

Another method of tailoring partitions in order to test for a specific prejudice, 
is to divide the supernova catalogue into a subset and a complementary set, respectively, below and above selected redshifts.
The motivation to do so is, for example, the shift of the supernova light-curves from visible bands to UV, e.g. around a redshift of $z=0.8$~\citep{Astier:2005qq},
which could systematically change the measurements, or as well the search for a signal of inhomogeneous cosmology.
In \cite{Amanullah:2010vv} the Union Compilation was already tested for redshift-dependent effects by forming 5 redshift bins and fitting stretch and colour correction as well as absolute magnitude in each bin, however fixing the cosmology. 
We take a different approach here and always fit the cosmology -- possibly using also the systematic parametrisation for the smaller subset when necessary -- in a fully Bayesian context.
As the division in redshift performed here yields only one subset plus complementary set per analysis, 
one is able, as in Section \ref{hemispe}, to analyse a higher number of synthetic catalogues (1000).
As for the parametrization chosen, we use a cosmological one for the subset and complementary set when partitioning up to redshift $z=0.3$.
For partitions at higher redshift the cosmological parameter estimation for the high-redshift subset fails because the likelihood contours become too degenerate.
Therefore the phenomenological parametrization of supernova magnitudes of Eq.~(\ref{eq:phenparam}) is adopted in these cases.
A chi-square test was performed in order to estimate the number of systematic parameters that reasonably parametrize the apparent magnitudes. 
For partitions at redshifts higher than $z=0.3$, the number of free parameters required was estimated to be two. 

\begin{table}
\caption{
Significance in $\sigma$-units of the robustness value relative to the Union2.0 Compilation with respect to unbiased synthetic catalogues for various partitions according to redshift.
A label ``c'' or ``s'' next to the redshift value indicates which parametrization (cosmological or systematic) was used. See Section \ref{cosmoz} for more details.
}
\centering
\def\arraystretch{1.1}
\begin{tabular}{| c| c |}
\hline \hline
z & Significance  \\
\hline
0.04 c & 0.99$\sigma$ \\
0.05 c& 0.86$\sigma$ \\
0.06 c& 1.04$\sigma$  \\
0.07 c& 0.67$\sigma$   \\
0.08 c& 1.23$\sigma$  \\
0.09 c& 0.65$\sigma$ \\ 
0.1 c& 0.56$\sigma$  \\
 0.2 c& 0.36$\sigma$  \\
 0.3 c& 0.57$\sigma$ \\
  0.3 s& 0.24$\sigma$ \\
\hline \hline
\end{tabular}
\begin{tabular}{| c| c| }
\hline \hline
 z & Significance \\
\hline
 0.4 s& 0.66$\sigma$  \\
 0.5 s& 0.04$\sigma$ \\
 0.6 s& 1.06$\sigma$  \\
 0.7 s& 0.42$\sigma$ \\
 0.8 s& 0.02$\sigma$ \\
 0.9 s& 0.95$\sigma$ \\
 1.0 s& 0.81$\sigma$ \\
 1.1 s& 0.94$\sigma$\\
 1.2 s& 0.22$\sigma$ \\
  1.3 s& 0.26$\sigma$ \\
\hline \hline
\end{tabular}
\label{table:sigmaz}
\end{table}

\begin{figure*}
\begin{center}
\includegraphics[width=0.95 \textwidth]{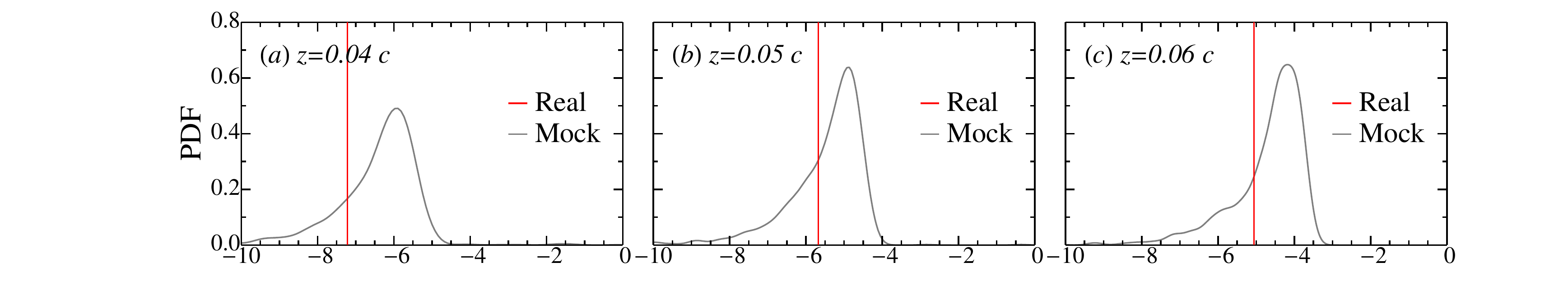}
\includegraphics[width=0.95 \textwidth]{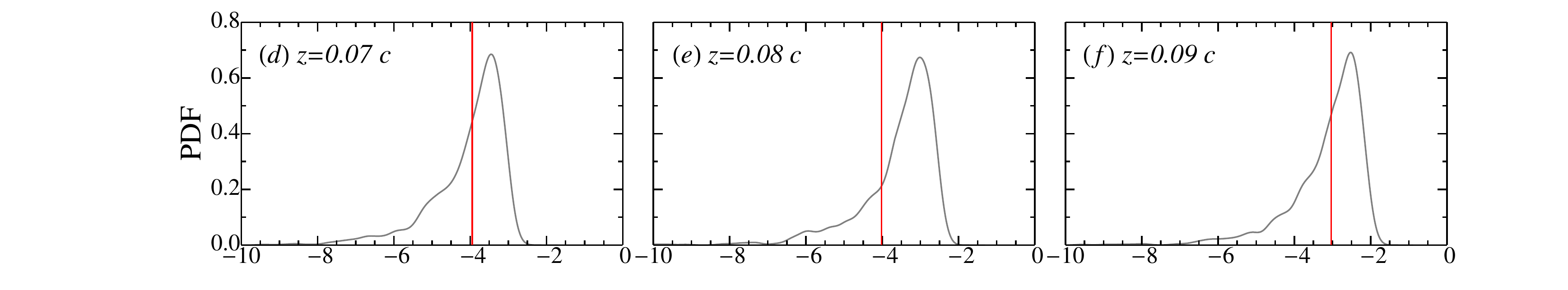}
\includegraphics[width=0.95 \textwidth]{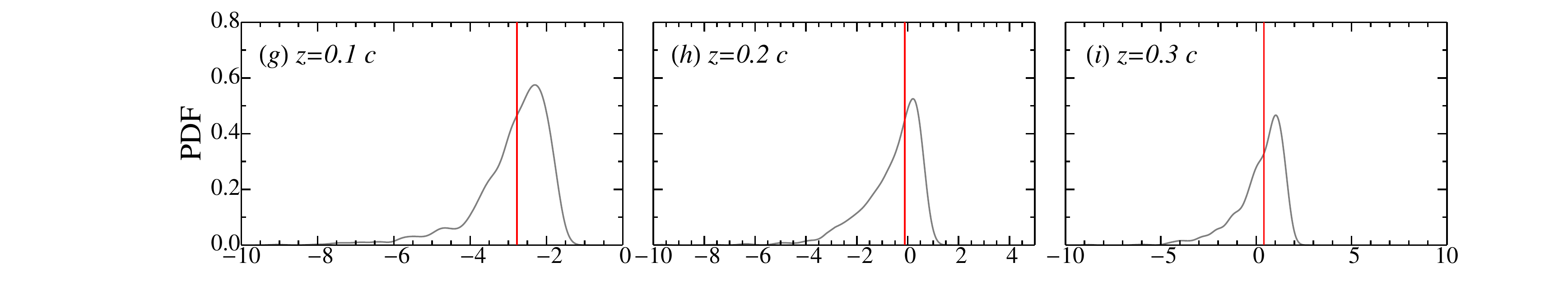}
\includegraphics[width=0.95 \textwidth]{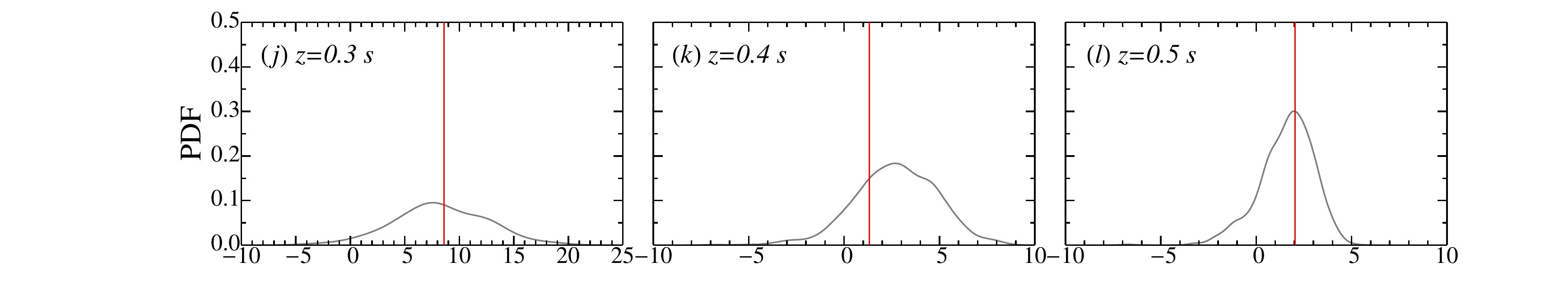}
\includegraphics[width=0.95 \textwidth]{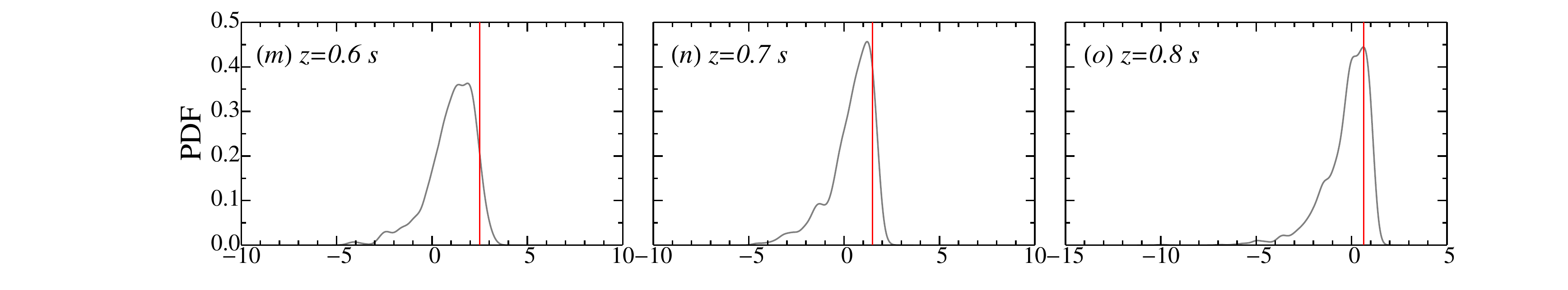}
\includegraphics[width=0.95 \textwidth]{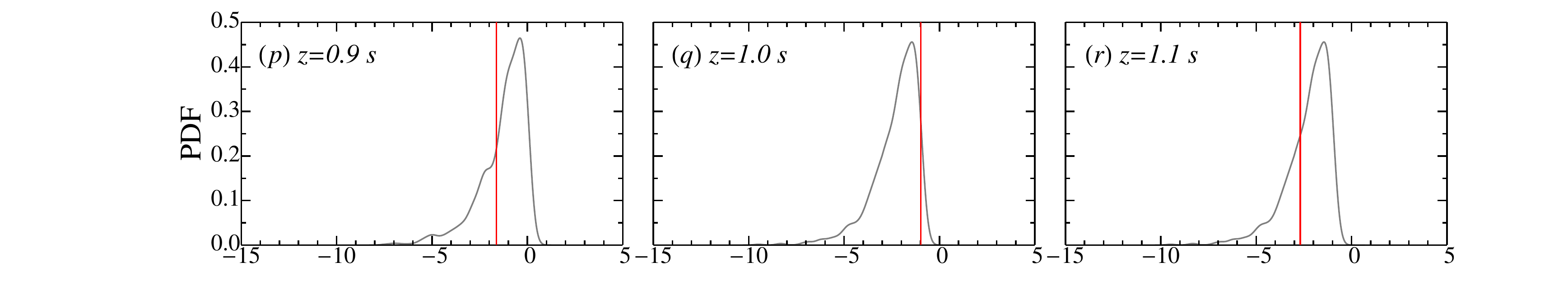}
\includegraphics[width=0.95 \textwidth]{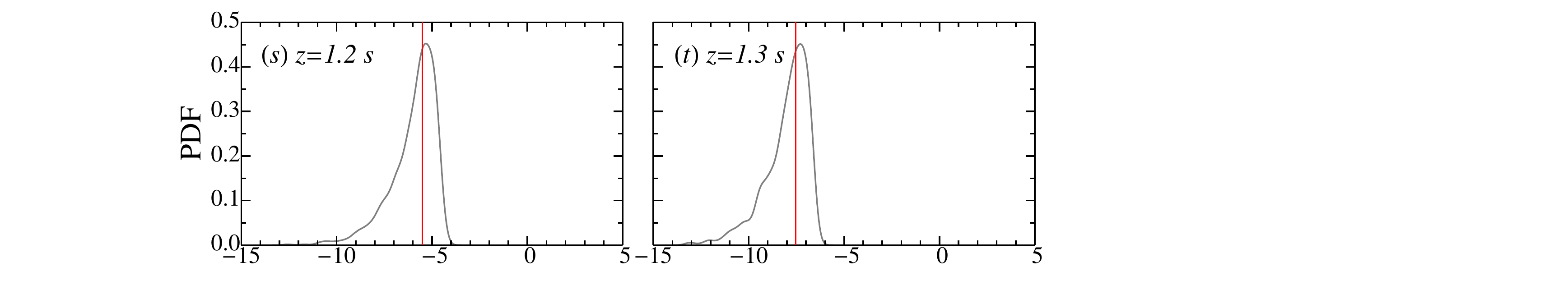}
\caption{
Robustness values relative to the Union2.0 Compilation (red vertical lines) and distributions from 1000 unbiased synthetic catalogues (grey solid lines) for various partitions according to redshift.
A label ``c'' or ``s'' next to the redshift value indicates which parametrization (cosmological or systematic) was used. 
The (low) significances of the signals are listed in Table \ref{table:sigmaz}.
See Section \ref{cosmoz} for more details.
}
\label{FIG:zlow}
\end{center}
\end{figure*}

We show the results of our analysis of the Union2.0 catalogue in Fig.~\ref{FIG:zlow}, for the indicated redshifts (used for partitioning). A label ``c'' or ``s'' next to the redshift value indicates which parametrization -- cosmological or systematic -- was used.
A minimum redshift $z=0.04$ was chosen so that the size of the smaller low-redshift subset would not be too small and cause the robustness evaluation to fail.
The corresponding significances of possible low-robustness signals (the red vertical lines in Fig.~\ref{FIG:zlow}) are listed in Table~\ref{table:sigmaz}.
As can be seen, the significance is always very low and does not display any clear trend with redshift,
therefore proving that the Union2.0 Compilation is robust 
against possible redshift-dependent systematical effects.

\subsection{Robustness of surveys} \label{surv}

The Union2.1 Compilation presented in~\cite{Suzuki:2011hu} comprises 19 different surveys,
each one with its own peculiarities and systematics.
The consistency of the Union2.1 Compilation as far as the different surveys are concerned has been already studied in \cite{Suzuki:2011hu} by comparing the average deviation of the different samples from the overall best-fit model.
Here we do not want to compare to an overall best-fit, potentially hiding/missing information, but directly assess if deviations of single surveys with respect to the other surveys are statistically detectable.

The various surveys differ in angular and redshift ranges covered and number of supernovae detected, as can be seen in Table~\ref{table:surprop} where the main properties of the 19 surveys are summarized.
In the present paper our aim is not to go into the details of the systematics of the various surveys and the way the latter were compiled consistently into one single catalogue, 
but rather to have an independent test of the robustness of each survey against the other surveys taken together.
The idea is to search for possible systematic effects hidden in the Union2.1 catalogue, which could potentially  influence parameter estimation.

\vspace{0.3cm}
\begin{table*}
\begin{minipage}{\textwidth}
\centering
\captionof{table}{Properties of the 19 different surveys making up the Union2.1 Compilation of~\citealt{Suzuki:2011hu} (first four columns).
Significance in $\sigma$-units of the robustness value relative to the Union2.1 Compilation with respect to unbiased synthetic catalogues when partitioning data according to each survey in turn, and using systematic and cosmological parametrization (fifth and sixth columns).
In the fifth columns, numbers in square brackets indicate the number of systematic parameters $\lambda_{i}$ used.
In the seventh (last) column a low-redshift SN sample was added to the survey being analysed, in case the latter had a degenerate likelihood.
When given in round brackets, the significance was obtained via a model parametrisation that actually showed degeneracies and/or
failure in chi-square testing.
See Section \ref{surv} for more details.
}
\begin{tabular}{| l | c | c | c | c | c | c | c| }
\hline 
 Survey No. & No. of SNe  & $z_{\rm min}$ &  $z_{\rm max}$ 
 &
\begin{minipage}[c][1.1cm]
[c]{2.7 cm}
\begin{center}
Significance
(systematic parametrization)
\end{center}
\end{minipage}  
&
\begin{minipage}[c][1.1cm]
[c]{2.7 cm}
\begin{center}
Significance
(cosmological parametrization)
\end{center}
\end{minipage}  
&
\begin{minipage}[c][1.1cm]
[c]{2.7 cm}
\begin{center}
Significance
(cosmological par. with low-$z$ sample)
\end{center}
\end{minipage}  
\\
\hline 
1 \citep{Hamuy:1996ss} & 18 & 0.0172 & 0.1009  &  0.94$\sigma$  [1]  &  (0.71$\sigma$) & - \\ 
2 \citep{Krisciunas:2005dw} & 6 & 0.0154 & 0.0305   &  0.16$\sigma$  [1] & - & - \\ 
3 \citep{Riess:1998dv} & 11 & 0.0152 & 0.1244  &  1.03$\sigma$  [2] & (0.17$\sigma$) & - \\ 
4 \citep{Jha:2005jg} & 15 & 0.0164 & 0.0537   &  1.04$\sigma$  [2] & - & - \\ 
5 \citep{Kowalski:2008ez} & 8 & 0.015 & 0.1561  &  0.04$\sigma$  [1]  & - & - \\ 
6 \citep{Hicken:2009dk} & 94 & 0.015 & 0.0843 &  0.26$\sigma$  [2]  & (1.94$\sigma$) & - \\ 
7 \citep{Contreras:2009nt} & 18 & 0.015 & 0.08  &- &- &- \\ 
8 \citep{Holtzman:2008zz} & 129 & 0.0437 & 0.4209 & (0.99$\sigma$  [3]) & 0.56$\sigma$ & - \\ 
9 \citep{Schmidt:1998ys} & 11 & 0.24 & 0.97  &  0.50$\sigma$  [1] &  (0.15$\sigma$) & - \\ 
10 \citep{Perlmutter:1998np} & 33 & 0.172 & 0.83  &  0.33$\sigma$  [3] & (0.42$\sigma$) & - \\ 
11 \citep{Barris:2003dq} & 19 & 0.3396 & 0.978  &-  &- &-  \\ 
12 \citep{Amanullah:2007yv} & 5 & 0.178 & 0.269 &-  &- & 0.02$\sigma$ \\ 
13 \citep{Knop:2003iy} & 11 & 0.355 & 0.86  &  0.03$\sigma$  [2] & - & 0.11$\sigma$ \\ 
14 \citep{Astier:2005qq} & 72 & 0.2486 & 1.01  &  (0.08$\sigma$  [2]) & 0.41$\sigma$ & 0.40$\sigma$ \\ 
15 \citep{Miknaitis:2007jd} & 74 & 0.159 & 0.781 &  (0.15$\sigma$  [3]) & 0.03$\sigma$ & 0.21$\sigma$ \\ 
16 \citep{Tonry:2003zg} & 6 & 0.278 & 1.057  & - &  - & 0.66$\sigma$  \\ 
17 \citep{Riess:2006fw} & 30 & 0.216 & 1.39  &  1.16$\sigma$  [2] & (2.11$\sigma$) & 0.73$\sigma$ \\ 
18 \citep{Amanullah:2010vv} & 6 & 0.511 & 1.124 & - &  - & 0.23$\sigma$ \\ 
19 \citep{Suzuki:2011hu} & 14 & 0.623 & 1.414  &  0.01$\sigma$  [1] & - & 1.53$\sigma$ \\ 
\hline 
\end{tabular}
\label{table:surprop}
\end{minipage}
\end{table*}
\vspace{0.3cm}

\begin{figure*}
\begin{center}
\includegraphics[width= .9 \textwidth]{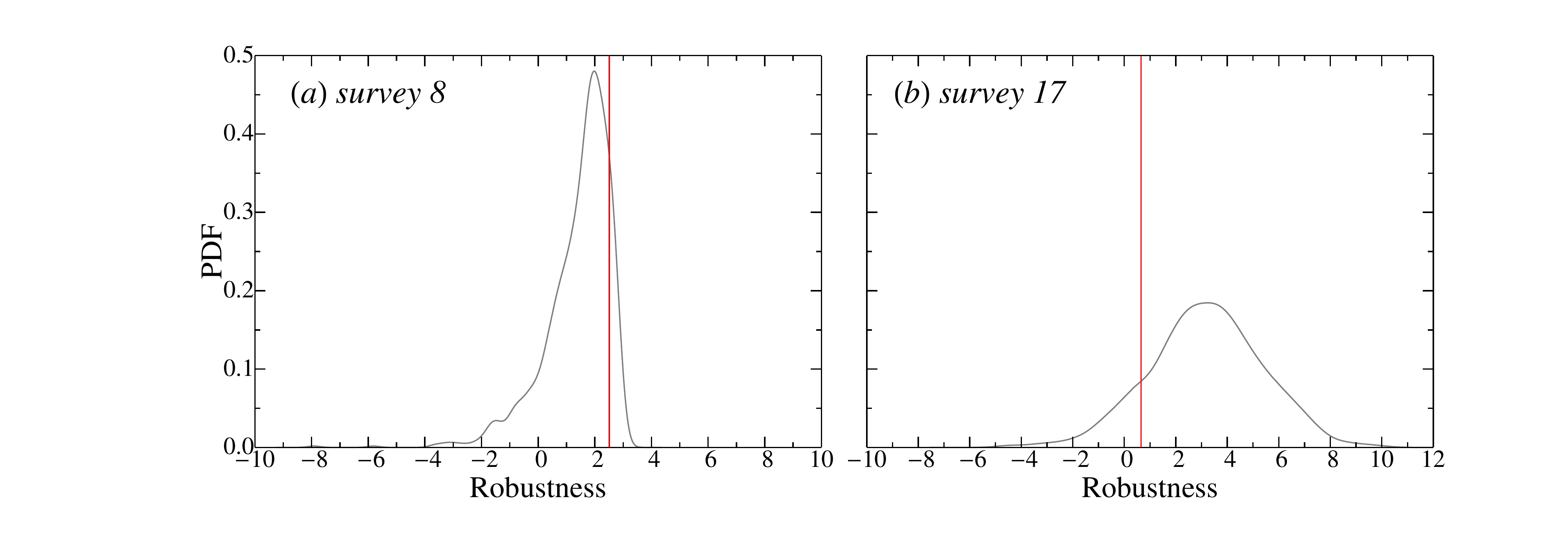}
\caption{
Left panel: survey No.~8 tested against all other surveys using the cosmological parametrization. The robustness value for the real catalogue is shown in red; the distribution of 1000 synthetic catalogues in grey. Right panel: same for survey No.~17.
The (low) significances of the signals are given in Table \ref{table:surprop}.
See Section \ref{surv} for more details.
}
\label{FIG:surv8}
\end{center}
\end{figure*}

As before we will calculate the internal robustness values for the real catalogue,
with each survey being in turn the (smaller) subset,
as well as the distribution of robustness values from 1000 synthetic unbiased catalogues.
We show in Fig.~\ref{FIG:surv8} the results for surveys No.~8 and 17.
Survey 8, together with surveys 14 and 15, was analysed using the cosmological parametrization. This is possible because these surveys contain a sufficient number of supernovae spread on a sufficiently large redshift range--condition necessary for having a non degenerate likelihood.
In order to test with the cosmological parametrization the robustness of surveys with otherwise degenerate likelihoods, we chose to add to the latter surveys the set of supernovae with $z<0.1$.
The significance of the corresponding values of internal robustness are given in Gaussian $\sigma$-units in Table~\ref{table:surprop}.

Survey 17 (right panel in Fig.~\ref{FIG:surv8}) was instead analysed employing the systematic parametrization of Eq.~(\ref{eq:phenparam}).
The appropriate number of parameters $\lambda_{i}$ to be used (in square brackets in Table~\ref{table:surprop}) was again found by performing a chi-square test for each survey individually.
The significances of the values of internal robustness for the surveys analysed with the systematic parametrization are again listed in Table~\ref{table:surprop}.

Summarizing the finding of this Section, neither survey No.~19 with the highest-redshift supernovae 
nor survey No.~1 being the oldest part of the compilation show a significant signal of systematics.
We conclude that the different surveys have been combined in quite a robust way.

\section{Conclusions}
\label{sec:conco}
In this paper we apply an advanced Bayesian statistical tool -- $\textit{internal robustness}$ --
to recent compilations of SNIa data, the Union 2.0 and 2.1 catalogues. 
Our aim is to quantify the presence of both systematic effects and cosmological signals unaccounted
for in previous analyses of the dataset.
Internal robustness enables us to search for subsets favoring a different underlying model than the
overall set,
without having to assume specific effects, and making at the same time use of all information
available in the full likelihood.
Our findings confirm a successful removal of systematics from the Union 2.0 and 2.1
compilations~\citep{Amanullah:2010vv,Suzuki:2011hu}, leaving only a low  level of systematics and proving these compilations most suitable
for cosmological parameter estimation. Furthermore, signals of anisotropy or inhomogeneity do not
seem to be significantly  reflected in the data.

Facing a huge number of possible partitions and striving to maximize our chances of finding the most
likely contaminated subsets,
we sorted the data by a variety of criteria: angular separation between pairs of supernova,
redshift, hemispheres on the celestial sphere and surveys that are a subset of the overall
compilation.

The analysis of the angular-separation-sorted supernovae shows no significant detection of
deviations, with highest signals being still within $2\sigma$.
The compilation thus is robust towards angular-separation-dependent effects.
The robustness of the compilation depending on redshift turns out to be at least as good, even at
high redshift,
proving successful removal of systematics-driven supernovae.
As regards our tests of celestial hemispheres, the anisotropies as reported by
Planck~\citep{Ade:2013nlj,Aghanim:2013suk} are not reflected in the SNIa data.
The direction of minimal robustness as found for the Union2.1 compilation corresponds to
$(\alpha,\delta)=(150^\circ,70^\circ)$. 
This also does not coincide with the directions of maximal anisotropy reported
in~\cite{Colin:2010ds,Kalus:2012zu,Cai:2013lja,Yang:2013gea,Rathaus:2013ut}.
The  relatively low level  of evidence for deviation from isotropy ($2.2 \sigma$) agrees with
earlier findings.
Other data than SN Ia could prove more appropriate to detect anisotropic signals.
Finally, the compilation of the 19 different surveys constituting Union 2.1 does not display a
significant signal of systematics
and therefore attests a robust combination of the different surveys.

Concluding, we can claim that the Union compilations have proven their robustness via this
independent cross-check,
even when sorting them in a way to maximize the incidence of a signal for both systematics and new
cosmology. 
An interesting future development could be extracting the most likely biased subsets of supernovae
having lowest internal robustness values.
Also very interesting could be to subdivide the supernova sample according to supernova and host galaxy observational properties, such as the host galaxy type and mass. Internal robustness could indeed help in confirming known correlations and finding new systematic effects.

\section*{Acknowledgements}

It is a pleasure to thank
Matthias Bartelmann, Emer Brady, Santiago Casas, Alexandre Posada, Miguel Quartin
for useful comments and discussions, and Ulrich Feindt, Marek Kowalski for sharing supernova data.
The authors acknowledge funding from DFG through the project TRR33 ``The Dark Universe''.
The Dark Cosmology Centre is funded by the DNRF. Finally, the authors want to thank the referee for useful comments and suggestions.

\bibliographystyle{mn2e_eprint}
\bibliography{references}



\bsp

\label{lastpage}

\end{document}